\documentclass[aps,prl,twocolumn,groupedaddress,superscriptaddress,showpacs]{revtex4-1}

\usepackage{graphicx}
\usepackage{color}
\usepackage{amsmath}
\usepackage{amsfonts}
\usepackage{amssymb}
\usepackage{dcolumn}

\begin{document}
  \title{Watersheds are Schramm-Loewner Evolution curves}

 \author{E. Daryaei}
 \email{daryaei@physics.sharif.edu}
 \affiliation{Computational Physics for Engineering Materials, IfB, ETH Zurich, Schafmattstrasse 6, 8093 Zurich, Switzerland}
 \affiliation{Department of Physics, Sharif University of Technology, P.O. Box 11155-9161, Tehran, Iran}

\author{N. A. M. Ara\'ujo}
  \email{nuno@ethz.ch}
  \affiliation{Computational Physics for Engineering Materials, IfB, ETH Zurich, Schafmattstrasse 6, 8093 Zurich, Switzerland}

\author{K. J. Schrenk}
  \email{jschrenk@ethz.ch}
  \affiliation{Computational Physics for Engineering Materials, IfB, ETH Zurich, Schafmattstrasse 6, 8093 Zurich, Switzerland}

\author{S. Rouhani}
 \email{srouhani@sharif.edu}
 \affiliation{Department of Physics, Sharif University of Technology, P.O. Box 11155-9161, Tehran, Iran}

\author{H. J. Herrmann}
  \email{hans@ifb.baug.ethz.ch}
  \affiliation{Computational Physics for Engineering Materials, IfB, ETH Zurich, Schafmattstrasse 6, 8093 Zurich, Switzerland}
  \affiliation{Departamento de F\'isica, Universidade Federal do Cear\'a, 60451-970 Fortaleza, Cear\'a, Brazil}


  \begin{abstract}
We show that in the continuum limit watersheds dividing
drainage basins are Schramm-Loewner
Evolution (SLE) curves, being described by one single parameter $\kappa$. 
Several numerical evaluations are
applied to ascertain this. All calculations are consistent with
SLE$_\kappa$, with $\mbox{$\kappa=1.734\pm0.005$}$, being the only known physical
example of an SLE with $\kappa<2$.
  This lies outside the well-known duality conjecture, 
bringing up new questions regarding the existence and
reversibility of dual models.
  Furthermore it constitutes a strong indication
for conformal invariance in random landscapes and suggests
that watersheds likely correspond to a logarithmic Conformal Field
Theory (CFT) with central
charge $\mbox{$c\approx-7/2$}$. 
 \end{abstract}

 \maketitle

The possibility of statistically describing the properties of random
curves with a single parameter fascinates physicists and mathematicians
alike. This capability is provided by the theory of Schramm-Loewner
Evolution (SLE), where random curves can be generated from a Brownian
motion with diffusivity $\kappa$ \cite{Schramm00}. Once
$\kappa$ is identified, several geometrical properties of the curve are
known (e.g.  fractal dimension, winding angle, and left-passage
probability) \cite{Cardy05,Bauer06}.  Among the examples of such
curves, we find self-avoiding walks \cite{Kennedy02} and the contours of
critical clusters in percolation \cite{Smirnov01}, $Q$-state Potts model
\cite{Rohde05}, and spin glasses \cite{Bernard07}, as well as in
turbulence \cite{Bernard06}. Establishing SLE for such
systems has provided valuable information on the underlying symmetries
and paved the way to some exact results
\cite{Lawler01,Smirnov01,Smirnov06}. In fact, SLE is not a general
property of non-self-crossing walks since many curves have been
shown not to be SLE as, for example, the interface of solid-on-solid
models \cite{Schwarz09}, the domain walls of bimodal spin glasses
\cite{Risau-Gusman08}, and the contours of negative-weight percolation
\cite{Norrenbrock12}.  

Recently, the watershed (WS) of random landscapes
\cite{Schrenk12,Cieplak94,Fehr11,*Fehr11b}, with a fractal
dimension $d_f\approx1.22$, was shown to be related to a family of
curves appearing in different contexts such as, e.g., polymers in
strongly disordered media \cite{Porto97}, bridge percolation
\cite{Schrenk12}, and optimal path cracks \cite{Andrade09}.  In the
present Letter, we show that this universal curve has the properties of SLE,
with $\kappa=1.734\pm0.005$.  $\kappa<2$ is a special limit since, up to
now, all known examples of SLE  found in Nature and statistical physics
models have $2\leq\kappa\leq8$, corresponding to
fractal dimensions $d_f$ between $1.25$ and $2$. 

Scale invariance and, consequently, the appearance of fractal dimensions
have always motivated to apply concepts from conformal invariance to
shed light on critical systems. Archetypes of self-similarity are the
contours of critical clusters in lattice models. Already back in 1923,
Loewner proposed an expression for the evolution of an analytic function
which conformally maps the region bounded by these curves into a standard
domain \cite{Loewner23}. Such an evolution, follows the theory, should
only depend on a continuous function of a real parameter, known as
\textit{driving function}. Recently,
Schramm argued that to guarantee conformal invariance, and domain Markov
property, the continuous function needs to be a one-dimensional Brownian
motion \cite{Schramm00} thrusting into motion numerous studies in what is
today known as the Schramm-Loewner, or Stochastic-Loewner,
Evolution (SLE).  Such one-dimensional Brownian motion has zero mean
value and is solely characterized by its diffusivity $\kappa$, which
relates with the fractal dimension $d_f$ as
\cite{Duplantier03,Beffara04},
\begin{equation}\label{eq::fractal.dimension}
d_f=\min\{1+\kappa/8,2\} \ \ .
\end{equation}
Although it is believed that SLE should hold for the entire class of
equilibrium O($n$) systems, it has only been rigorously proven for a few cases
\cite{Smirnov01,Smirnov06}.  Nevertheless, numerically 
correspondence has been shown for a large number of models as mentioned
above. It has been argued that SLE can be applied to models exhibiting
non-self-crossing paths on a lattice, showing self-similarity, not only in
equilibrium but also out of equilibrium as the example discussed here
\cite{Amoruso06,Bernard06,Saberi09}.

In random discretized landscapes each site is
characterized by a real number such as, e.g., the height in an
elevation map, the intensity in a pixelated image, or the energy in an
energy landscape \cite{Schrenk12}. If
sites are occupied from the lowest to the highest, clusters of adjacent occupied
sites can be defined and, at a certain fraction of occupied sites, a
spanning cluster emerges connecting opposite borders (e.g. from left to
right). In the example of the elevation map, this
procedure corresponds to filling the landscape with water until
a giant lake emerges at the threshold, which drains to the borders
\cite{Knecht11}. When we now suppress spanning by imposing the
constraint that sites, merging two clusters touching the opposite
borders, are never occupied, a line emerges
delineating the boundaries between two clusters: one connected to the
left and the other to the right border \cite{Schrenk12}. This line is
the watershed line (WS) separating two hydrological basins \cite{Fehr09}
and has a fractal dimension $d_f=1.2168\pm0.0005$ \cite{Fehr11c}.  We
show in this Letter that in the scaling limit its statistics converges to
an SLE$_\kappa$, consistent with $\kappa=1.734\pm0.005$.

To study the scaling limit of WS and compare it to SLE$_\kappa$, we
numerically generated ensembles of curves and carried out three
different statistical evaluations, namely, the variance of the \emph{winding
angle} (quantifying the angular distribution of the curves)
\cite{Duplantier88,Wieland03}, the \emph{left-passage probability}
\cite{Schramm00,Schramm01}, and the characterization of the driving function
(\emph{direct SLE}) \cite{Bernard06}. We show here that the values of
$\kappa$ independently obtained for each analysis are numerically consistent
and in line with the fractal dimension of the WS. For all cases,
simulations have been performed on both, square lattices of square shape
($L_x=L_y$) and in strip geometries ($L_x>L_y$), all with free boundary
conditions in horizontal and periodic boundary conditions in vertical
direction. $L_x$ is the size of the horizontal 
boundary, while $L_y$ is the length of the vertical one.
Hereafter, we discuss each analysis separately. 

\begin{figure}[t]
\begin{center}
\includegraphics[width=\columnwidth]{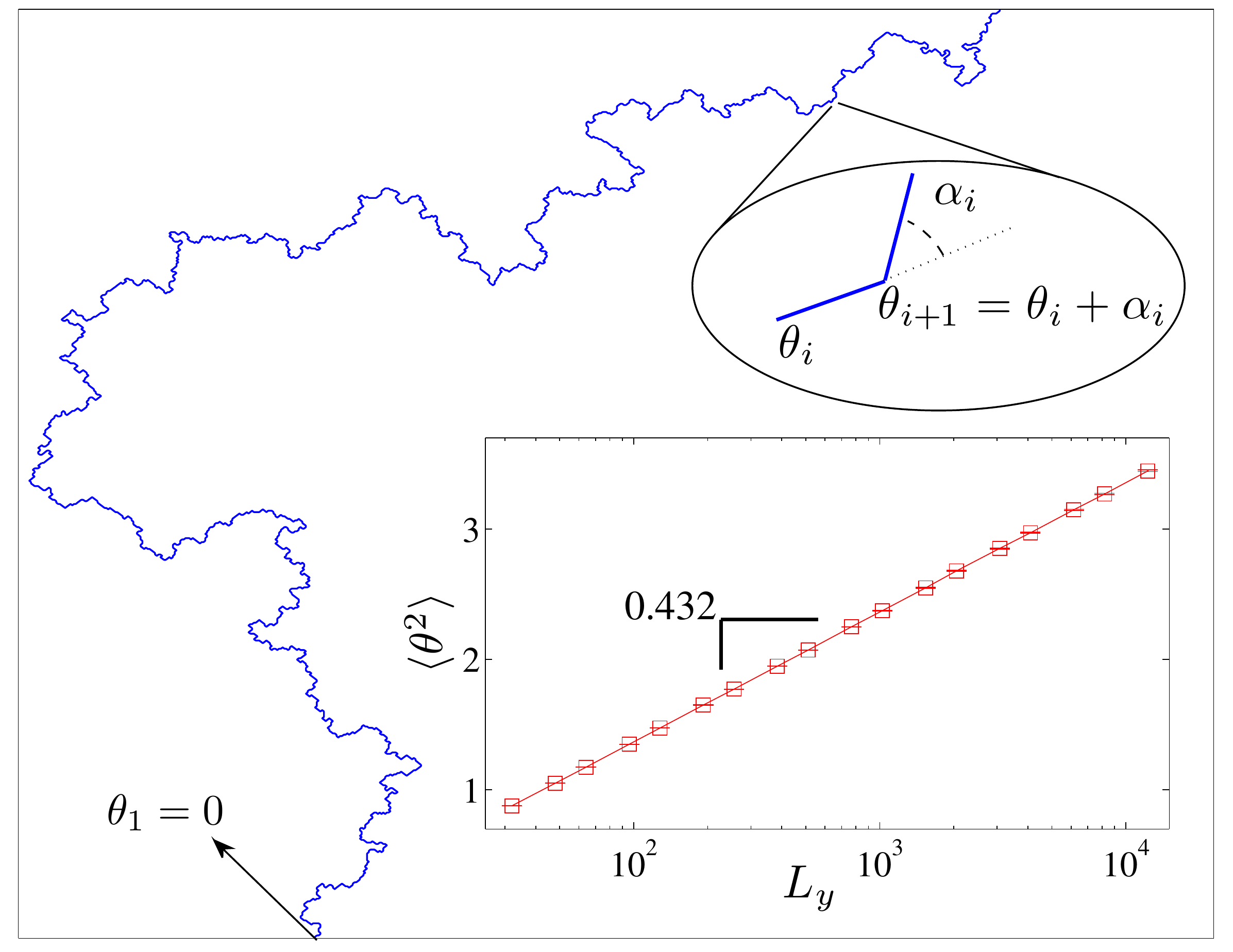}
\end{center}
\caption{
(color online) Defining $\alpha_i$ as the turning angle between two
adjacent edges, $i$ and $i+1$, their winding angles $\theta_i$ and
$\theta_{i+1}$ are related by $\theta_{i+1}=\theta_i+\alpha_i$, with
$\theta_1=0$, as
illustrated in the main figure. We performed
simulations for 18 different lattice sizes, with $L_y=2^{4+n}$ and $3
\times 2^{3+n}$ for $n=1,2,\dots,9$. Results are averages over $10^6$
samples for the smallest system sizes and $3\times10^3$ for the largest
one.  \textbf{Inset}: Dependence of the variance of the winding angle on
the lateral size of the lattice $L_y$. Statistical error bars are 
smaller than the symbols.  The slope in the linear-log plot corresponds
to $\kappa/4=0.432\pm0.002$. 
\label{fig::winding.angle}
}
\end{figure} 
\emph{Winding angle}.~Using conformal invariance and Coulomb-gas
techniques, Duplantier and Saleur \cite{Duplantier88} have found the
dependence of the distribution of the winding angle on the system size
and the Coulomb-gas parameter. Given the correspondence of the
Coulomb-gas parameter to $\kappa$, the relation for the winding angle
can be extended to SLE \cite{Wieland03}.  To analyze the winding angle
$\theta_i$ at edge $i$, we set $\theta_1=0$ and define $\alpha_i$ as the
turning angle between the edges $i$ and $i+1$ (see
Fig.~\ref{fig::winding.angle}). The winding angle of each edge is then
computed iteratively as $\theta_{i+1}=\theta_i+\alpha_i$. For
SLE$_\kappa$, the variance of the winding angle over all edges in the
curve scales as $\langle\theta^2\rangle=b+(\kappa/4)\ln L_y$, where $b$
is a constant and $L_y$ the lateral size of the lattice \footnote{There is
some discussion on the proper way to measure the winding angle
\cite{Wieland03}. In this work we follow the definition described in the
text.}.  Figure~\ref{fig::winding.angle} shows the variance as function
of lateral size $L_y$ for the WS, with a slope $0.432\pm0.002$ in a
linear-log plot. This slope corresponds to $\kappa=1.728\pm0.008$ which
is in good agreement with the one predicted by
Eq.~(\ref{eq::fractal.dimension}) from the WS fractal dimension.

\begin{figure}[t]
\includegraphics[width=\columnwidth]{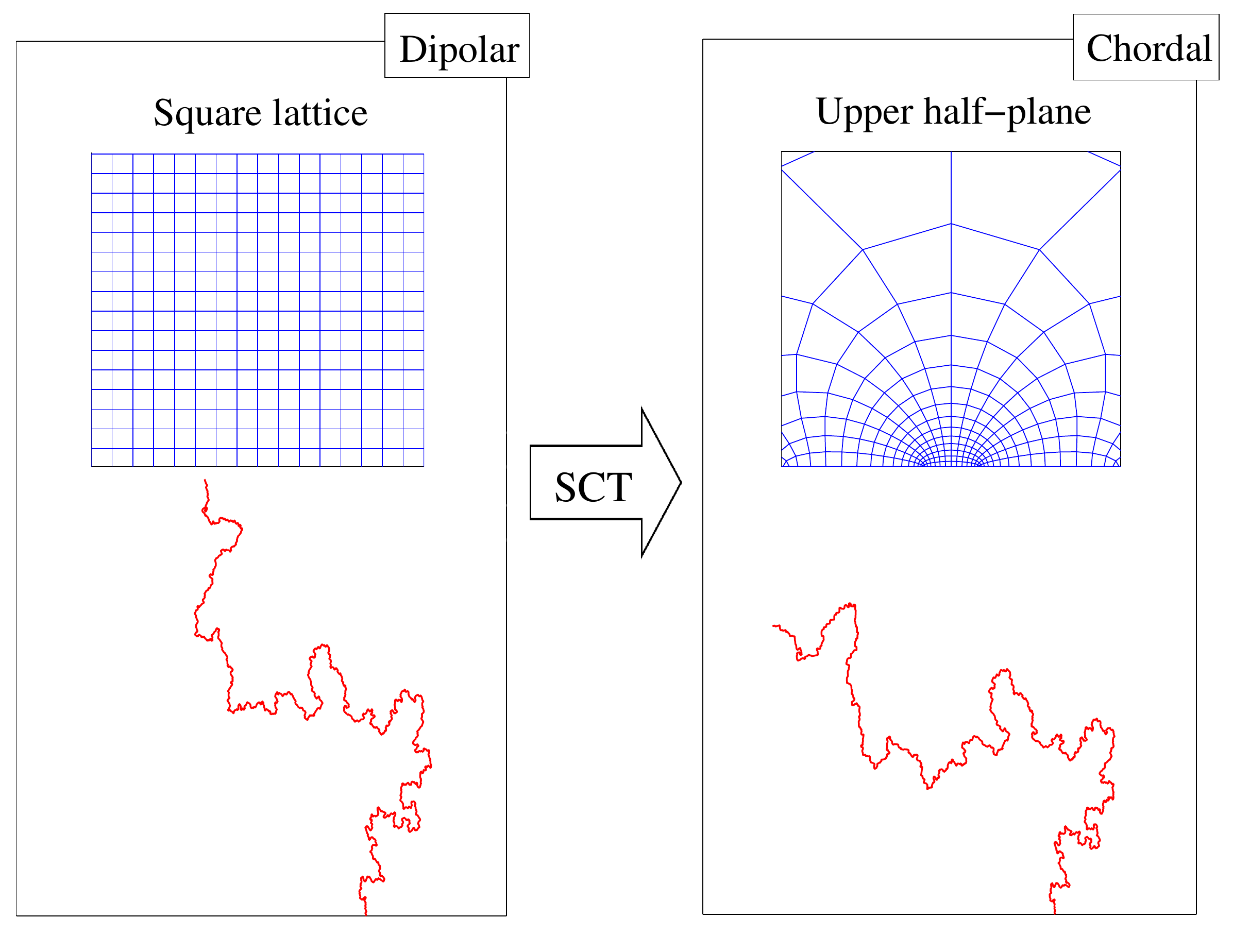}
\caption{
(color online) Original (left) and mapped (right) watershed. The Schwartz-Christoffel
transformation (SCT) has been applied to map from the square
lattice (left) to the upper half-plane (right). In this way dipolar
curves are turned into chordal curves.
\label{fig::sct}
}
\end{figure} 
\emph{From dipolar to chordal representation.}~In the original setup, WS
are \emph{dipolar} curves which start at 
one point on the lower boundary and end when they touch the upper
boundary, for the first time. For the left-passage probability and
direct SLE evaluations, exact
results are however known for \emph{chordal} curves
\cite{Schramm01}, which start at the same point but
go to infinity. Therefore, to proceed with these evaluations, we map the
dipolar WS curves into chordal ones in the upper half-plane $\mathbb{H}$
(see Fig.~\ref{fig::sct}). For such mapping, as suggested in
Refs.~\cite{Chatelain12,Driscoll96}, we used the inverse
Schwartz-Christoffel transformation \footnote{We used the algorithm
described in Ref.~\cite{Driscoll96}. Since with the Schwartz-Christoffel
transformation, the vertices of the square lattice are mapped to the
real axis in $\mathbb{H}$, to avoid the mapped curve to return to the
real axis, we mapped a square domain $[-1,1]\times[0,2]$ with the curve
constrained to the domain $[-0.5,0.5]\times[0,1]$.}.

\begin{figure}[t]
\includegraphics[width=\columnwidth]{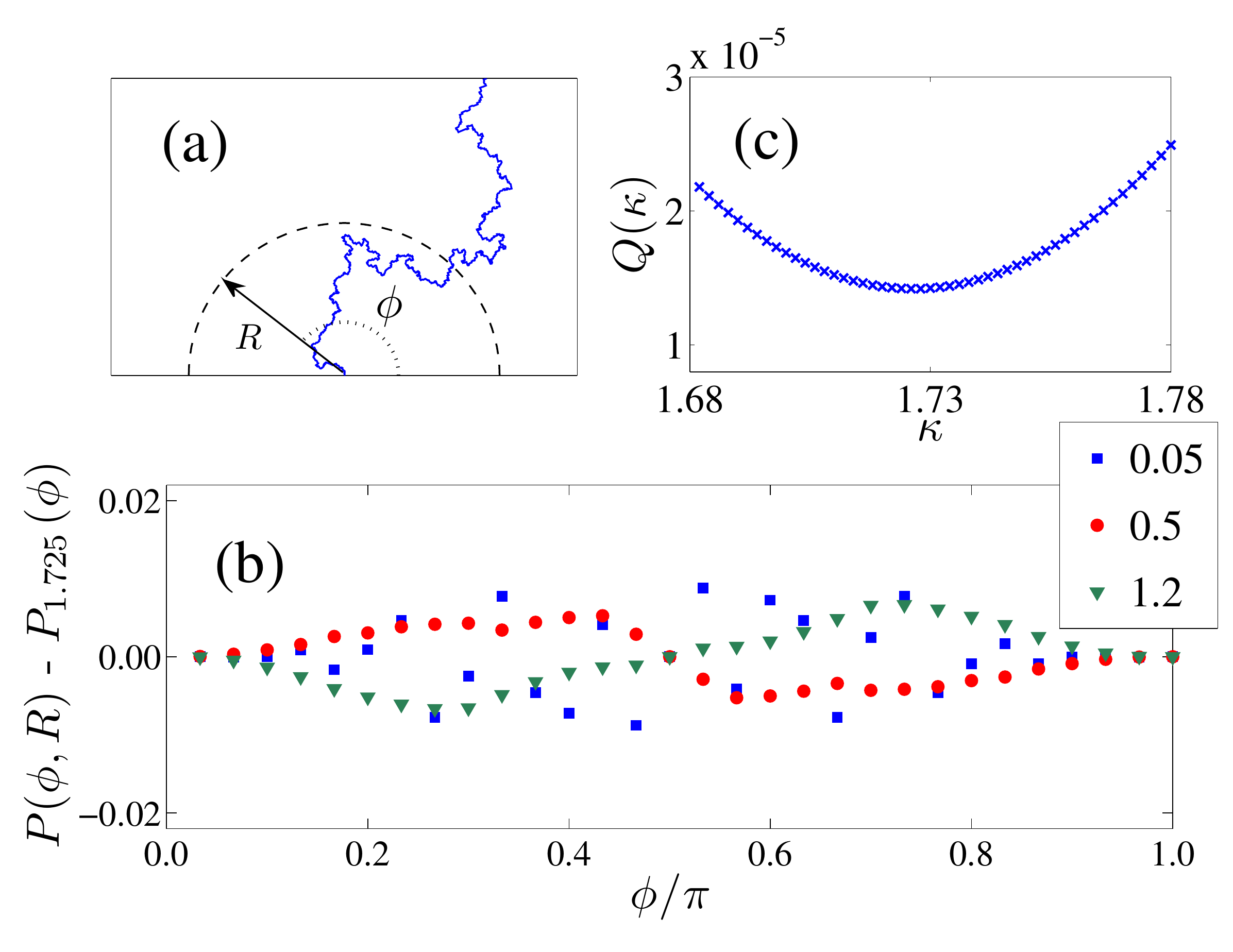}
\caption{
(color online) (a) Schematic representation of the left-passage
definition (details in the text). (b) $P(\phi,R)-P_{1.725}(\phi)$ for
the chordal watershed at different distances from the origin
$R=\{0.05,0.5,1.2\}$, where $P_{1.725}(\phi)$ is the left-passage
probability for $\kappa=1.725$ given by Schramm's formula,
Eq.~(\ref{eq::left.passage}). (c) Mean square
difference $Q(\kappa)$ between the numerical data and Schramm's
formula (Eq.~(\ref{eq::deviation})) for different values of
$\kappa$, exhibiting a minimum at $\kappa=1.73\pm0.01$. In both cases,
results are averages over $10^5$ curves on square lattices with $L_y=512$.
\label{fig::left.passage}
}
\end{figure}

\emph{Left-passage probability}.~For SLE curves in the upper half-plane
$\mathbb{H}$, starting at the origin, the probability that a point $R
e^{i\phi}$ is at the right side of the curve (see
Fig.~\ref{fig::left.passage}(a)) solely depends on $\phi$ and $\kappa$
and is given by Schramm's formula \cite{Schramm01},
\begin{equation}\label{eq::left.passage}
P_{\kappa}(\phi)=\frac{1}{2}+\frac{\Gamma\left(\frac{4}{\kappa}\right)}{\sqrt{\pi}
\Gamma\left(\frac{8-\kappa}{2\kappa}\right)}\cot(\phi)_2F_1
\left(\frac{1}{2};\frac{4}{\kappa},\frac{3}{2};-\cot^{2}(\phi)\right) \ \ ,
\end{equation}
where $_2F_1$ is the Gaussian hypergeometric function and $\Gamma$ is
the Gamma function. Figure~\ref{fig::left.passage}(b) are the data
points for the difference between the numerically measured probability
$P(\phi,R)$ and the one predicted by Schramm's formula,
Eq.~(\ref{eq::left.passage}), for the chordal curve. It is shown that
$P(\phi,R)$ is independent on $R$. To estimate
$\kappa$ we plot, in Fig.~\ref{fig::left.passage}(c), the mean square
deviation $Q(\kappa)$ defined as,
\begin{equation}\label{eq::deviation}
Q(\kappa)=\frac{1}{M}\sum_R\sum_\phi\left[P(\phi,R)-P_\kappa(\phi)\right]^2 \ \ ,
\end{equation}
where the outer sum goes over values of $0.05\leq R\leq 1.2$, in steps of
$0.05$, and the inner one over values of $0\leq\phi\leq\pi$, in steps
of $\pi/15$. $M$ is the total number of considered points $R e^{i\phi}$.
To reduce the statistical noise we used the relation
$P(\phi,R)+P(\pi-\phi,R)=1$. The minimum in the plot corresponds to the
value of $\kappa$ that best fits the left-passage probability, giving
\mbox{$\kappa=1.73\pm0.01$}, in line with the prediction based on the
fractal dimension of WS, given by Eq.~(\ref{eq::fractal.dimension}).

\emph{Direct SLE}.~Consider a chordal SLE curve $\gamma(t)$ which starts
at a point on the real axis and grows to infinity inside the region of
the upper half-plane $\mathbb{H}$, parametrized by an adimensional
parameter $t$, typically called Loewner time. To compute its driving
function $\xi(t)$ one needs to find the sequence of maps $g_t(z)$ which
at each time $t$ map the upper half-plane $\mathbb{H}$ into $\mathbb{H}$
itself and satisfy the Loewner equation \cite{Loewner23}. This map
is unique and can be approximately obtained by considering
the driving function to be constant within an interval $\delta t$, 
obtaining the slit map,
\begin{equation}\label{eq::slit.map}
g_t(z)=\xi(t)+\sqrt{\left(z-\xi(t)\right)^2+4\delta t} \ \ ,
\end{equation}
where $z$ is a point in $\mathbb{H}$ and $\delta t$ also depends on $t$.
This map converges to the exact one for vanishing $\delta t$
\cite{Bauer03}.  Initially, we set $t=0$ and $\xi(0)=0$ and we proceed
iteratively through all points $z_i$ of the chordal curve. At each
iteration $j$, we map the point $z_j$ to the real axis, by setting
$\delta t_j=\left(\text{Im } z_j\right)^2/4$ and the driving function
$\xi(t_j)=\text{Re } z_j$ (being Re and Im the real and imaginary parts,
respectively). We also compute the Loewner time $t_j=t_{j-1}+\delta
t_j$. As referred above, in SLE, the driving function is related to a
Brownian motion $B(t)$, with vanishing mean value and unit dispersion,
such that $\xi(t)=\sqrt{\kappa}B(t)$ \cite{Schramm01}.
\begin{figure}[t]
\includegraphics[width=\columnwidth]{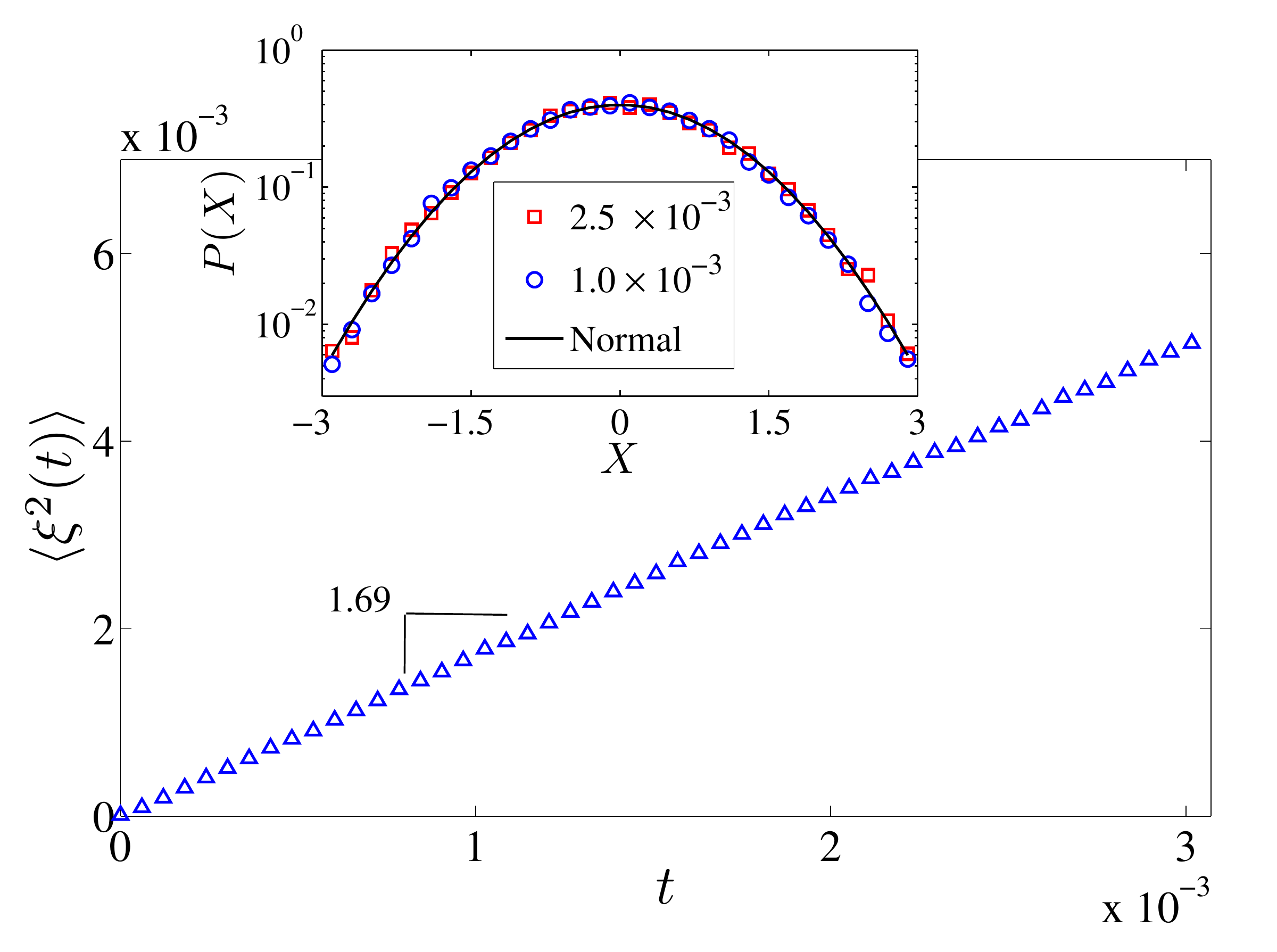}
\caption{
(color online) Dependence of the second moment of the driving function
$\langle\xi^2(t)\rangle$ on the Loewner time, for the chordal watershed.
The slope corresponds to $\kappa=1.69\pm0.05$.  \textbf{Inset}:
Probability distribution of the driving function at two different
Loewner times for chordal watersheds. The rescaled parameter $X$ is
defined as $X=\xi(t)/\sqrt{\kappa t}$, where we have taken
$\kappa=1.69$. The solid line is the normal distribution of vanishing
mean value and unit dispersion.  Results are averages over $4\times10^4$
realizations on a square lattice with $L_y=1024$.
\label{fig::direct.sle}
}
\end{figure} 

Figure~\ref{fig::direct.sle} shows the second moment of the driving
function for the chordal WS. The inset displays the probability
distribution for the rescaled driving function
\mbox{$X=\xi(t)/\sqrt{\kappa t}$} for two different times for the
chordal WS. All results are consistent with a Brownian motion with
vanishing mean value and unit dispersion, when $\kappa=1.69\pm0.05$, in good
agreement with the results discussed above. The direct SLE analysis is
characterized by larger error bars than the other two methods (winding
angle and left-passage probability) due to strong discretization effects
in the slit mapping \cite{Bauer03}.

\emph{Discussion}.~Our detailed numerical analysis shows that watersheds
are likely to be SLE curves with \mbox{$\kappa=1.734\pm0.005$}.  This is
the first documented case of a physical model with $\kappa<2$, lying
outside the well-known duality conjecture range $2\leq\kappa\leq8$,
giving $\kappa'=16/\kappa$, where $\kappa'$ is the diffusivity of the
dual model \cite{Dubedat05}. It has been proven that SLE$_\kappa$ with
$\kappa>8$ is not reversible \cite{Zhan08}, therefore if a dual model
exists which respects reversibility, then it cannot be SLE$_{\kappa'}$
with ${\kappa'}>8$. In the context of SLE, duality implies that
two apparently different fractal dimensions might actually stem from the
same curve. Geometrically, this corresponds to a relation between the
fractal dimension of the accessible external perimeter and the one of the
curve. 

Our work shows that watersheds are non-local SLE curves.
Although a connection with SLE is strong indication for conformal
invariance, it cannot be interpreted as a proof. Nevertheless, if
such invariance is established, it becomes possible to develop a field
theory for this new universality class. CFT has helped to classify
continuous critical behavior in two-dimensional equilibrium phenomena
\cite{DiFrancesco97,Henkel12}. 
 A well-established relation between diffusivity
$\kappa$ and central charge $c$ of minimal CFT models which have a
second level null vector in their Verma module is
$c=(3\kappa-8)(6-\kappa)/2\kappa$ \cite{Bauer03}. 
If the
watershed is conformally invariant it likely corresponds to a
logarithmic CFT (LCFT) with central charge $c\approx-7/2$. A series of
LCFT's corresponding to loop models have been suggested in
Ref.~\cite{Provencher11}, which thus seem to be related to watersheds.
It is also noteworthy
that negative central charges have been reported in different contexts
like, e.g., stochastic growth models, $2D$ turbulence, and quantum
gravity \cite{Duplantier92,Flohr96,Lipatov99}.  In particular, the loop
erased random walk is believed to have $\kappa=2$ which corresponds to
$c=-2$. 
Besides, since the watershed of
a landscape is based on the distribution of heights, the
configurational space grows with $N!$, where $N$ is the number of sites,
being a promising candidate to develop a
field theory with quenched disorder. 

The connection between SLE and statistical properties of the watershed
opens up new possibilities. Since the latter are related to fractal
curves emerging in several different contexts, our work paves the way to
bridge between connectivity in disordered media and optimization
problems where the same $\kappa$, and its corresponding central charge, are
observed. Besides, a systematic study of the $\kappa$ dependence on
correlations in the landscape might provide the required information to
find SLE curves on natural landscapes. The possibility of a multifractal
spectrum for watersheds is also an open question.

\begin{acknowledgments}
We acknowledge financial support from the ETH Risk Center.  We also
acknowledge the Brazilian institute INCT-SC. ED also acknowledges
financial support from Sharif University of Technology during his visit
to ETH as well as useful discussion with C. Chatelain, E. Dashti, and M.
Rajabpour. NAMA thanks J. P. Miller and B. Duplantier for some helpful
discussion.
\end{acknowledgments}

\bibliography{sle}
     
\end{document}